\documentclass{elsart}

\usepackage{epsfig,subfigure,latexsym,amsmath}

\begin{document}

\baselineskip 0.6cm

\def\Journal#1#2#3#4{{#1} {\bf #2}, #3 (#4)}
\def\RPP{{Rep.Prog.Phys.}}
\def\PRC{{Phys. Rev. C.}}
\def\FP{{Foundations of Physics}}
\def\ZPA{{Z. Phys.A.}}
\def\NPA{{Nucl. Phys. A.}}
\def\JPG{{J. Phys. G Nucl. Part}}
\def\PRL{{Phys. Rev. Letts.}}
\def\PRpt{{Phys. Report.}}
\def\PLB{{Phys.Letts. B}}
\def\AP{{Ann. Phys (N.Y.)}}
\def\EPJA{{Eur. Phys. J.A}}
\def\NP{{Nucl. Phys}}
\def\ZP{{Z. Phys}}
\def\RMP{{Rev. Mod. Phys}}

\input epsf

\begin{frontmatter}

\title{The nonrelativistic limit of the relativistic point coupling model}

\author[indo]{A. Sulaksono}
\author[lanl]{T. B{\"u}rvenich}
\author[fra]{J.~A. Maruhn}
\author[er]{P.--G. Reinhard}
\author[fra]{W. Greiner}
\address[indo]{Jurusan fisika, FMIPA, Universitas Indonesia, Depok 16424, Indonesia}
\address[lanl]{Theoretical Division, Los Alamos National Laboratory, Los Alamos, New Mexico 87545, USA}
\address[fra]{Institut f\"ur Theoretische Physik, Universit\"at
Frankfurt, Robert-Mayer-Str. 10, D-60324 Frankfurt, Germany}
\address[er]{Institut
f\"ur Theoretische Physik II, Universit\"at Erlangen-N\"urnberg, Staudtstrasse 7, D-91058 Erlangen, Germany}

\begin{abstract}
We relate the relativistic finite range mean-field model (RMF-FR) 
to the point-coupling variant and compare the nonlinear density dependence.
From this, the effective Hamiltonian of the nonlinear
point-coupling model in the nonrelativistic limit is derived.
Different from the nonrelativistic models,
the nonlinearity in the relativistic models automatically yields
contributions in the form of a weak density dependence not only in the central
potential but also in the spin-orbit potential. The central potential
affects the  bulk and surface properties while the
spin-orbit potential is crucial for the shell structure of 
finite nuclei. A modification in the Skyrme-Hartree-Fock model with a
density-dependent spin-orbit potential inspired by the point-coupling model
is suggested.
\end{abstract}

\begin{keyword}
Skyrme-Hartree-Fock model \sep relativistic mean-field model  \sep nonrelativistic limit
\PACS 
21.30.Fe, 21.60.Jz
\end{keyword}
\end{frontmatter}

\section{Introduction}
Relativistic point-coupling (RMF-PC) models have proven to deliver
predictions for nuclear ground-state observables which are of
comparable quality as the ones from the well-established finite-range
relativistic mean-field (RMF-FR) model (for a review see \cite{pg})
and the Skyrme-Hartree-Fock (SHF) approach (for a review see
\cite{que78}) \cite{niko,buer01}. Besides opening up the way to
relativistic Hartree-Fock calculations (see Ref. \cite{anto2} for a recent application) numerically similar to Hartree
calculations with the use of Fierz relations for the exchange terms
(up to fourth order, see Ref. \cite{ma}) and to the study of the role
of {\em naturalness} \cite{fri96,mg84} in effective theories for
nuclear structure related problems, the RMF-PC approach also provides
an opportunity to study the interrelations between nonrelativistic and relativistic point-coupling models, i.e., between the RMF-PC and the SHF approach.  In
Ref.~\cite{sur} Rusnak and Furnstahl have shown the
profitability to apply the concepts of effective field theory such as
naturalness to 
point-coupling models, where besides the version which will be used
in our analysis, they
consider also tensor terms, the mixing among the densities in
the nonlinear terms, and nonlinear derivative terms. This
investigation is motivated by the fact that up till now the role and
the importance of the various terms in the  RMF-PC ansatz is not completely understood, not speaking about terms which might yet be missing.
There appear systematic differences in the model predictions which 
could not yet be mapped onto the corresponding features of the models.
A comparison between the nonrelativistic limit of the RMF-PC model and
the SHF model may help to clarify these questions.

One problem to face when comparing relativistic finite range with
point-coupling nonrelativistic models is that two limits have to be
taken:  (a) the limit of letting the range of the mesons shrink to zero and (b) the expansion in powers of $v/c$ (nonrelativistic reduction).
The connection between the Skyrme-Hartree-Fock  model and the RMF-FR model  was done by several
authors, without~\cite{the,hor} and with
nonlinear terms~\cite{pg,abou} (employed as self-interactions of the $\sigma$-meson), but they did not take into account tensor
contributions. The role of the tensor coupling of the isoscalar 
vector meson to the nucleon in the framework of effective field theories
was investigated in Ref.~\cite{furn1}.
 
The nonlinear density dependence is introduced in much different
fashion for Skyrme-Hartree-Fock as compared to the Walecka model which
employs non-linear meson self-couplings. This is a hindrance for a
direct comparison \cite{pg,abou}. On the other hand, nonlinear terms
in RMF models are important, because only relativistic models with
nonlinear terms can reproduce experimental data with acceptable
accuracy~\cite{pg,rufa,niko}. We avoid the problems if we use the
RMF-PC model~\cite{niko}, because in this model the nonlinear terms
are explicitly density dependent, similar as in Skyrme-Hartree-Fock.
Therefore it is worthwhile to study the connection between
RMF-FR and RMF-PC on one hand and between the RMF-PC model and the
nonrelativistic Skyrme-Hartree-Fock model on the other hand. In this
context, we can study the role not only of the linear terms but also
the nonlinear ones of both models in the nonrelativistic limit. 

The paper is outlined as follows: Section \ref{FRtoPC} evaluates the
expansion of the finite range meson propagators of the RMF-FR into
point-couplings.  The non-relativistic limit is then discussed in
section \ref{PCtoSHF} from the RMF-PC as starting point. The part
discussing in particular the emerging structure of the spin-orbit
functional is taken up in section \ref{densdepls}. And a few general
comments on exchange are finally made in section \ref{ex}.

\section{From RMF-FR to RMF-PC}
\label{FRtoPC}
The RMF-PC model can be considered as the mediator between RMF-FR and
SHF. The effects of finite range have nothing to do with the
nonrelativistic limit, so we study them first by comparing the
zero-range limit of RMF-FR with the point-coupling ansatz while
remaining at the level of RMF. Having done this, we proceed in the
subsequent section with the derivation of the nonrelativistic limit of
the RMF-PC model.

The covariant formulation of the RMF is based on a 
Lagrangian density. It is given for the RMF-FR in
appendix \ref{app:RMF} . For the stationary case, it can equally well be 
formulated as a Hamiltonian density. This reads
\begin{subequations}
\begin{eqnarray}
   {\mathcal H} 
   &=& 
   {\mathcal H}_{\rm free}^{\rm nuc}
   + {\mathcal H}_{\rm S}+ {\mathcal H}_{\rm V} +{\mathcal H}_{\rm R},
\\
   {\mathcal H}_{\rm free}^{\rm nuc} 
   &=& 
   \sum_{\alpha} \bar{\Psi}_{\alpha} ( - i
   \vec{\gamma}. \vec{\nabla} +m_B ) \Psi_{\alpha},
\\
   {\mathcal H}_{\rm S} 
   &=& 
   \frac{1}{2}\left((\nabla\Phi)^2+m_S^2\Phi^2\right)
   +g_{\rm S} \Phi \rho_{\rm S}+\frac{1}{3}
   b_2 \Phi^3+\frac{1}{4} b_3 \Phi^4,
\\
   {\mathcal H}_{\rm V} 
   &=&  
   -\frac{1}{2}\left(\nabla V^\mu\nabla V_\mu+m_V^2V^\mu V_\mu\right)
   +g_{\rm V}\rho_0 V_0 - \frac{f_{\rm V}}{2 m_B}\rho_{\rm T}V_0, 
\label{eq:HinitV}\\
  {\mathcal H}_{\rm R} 
  &=&  
   -\frac{1}{2}\left(\nabla R^\mu_\tau\nabla R_{\mu,\tau}
   +m_V^2R^\mu_\tau R_{\mu,\tau}\right)
  +g_{\rm R}\rho_{\rm \tau 0} R_{\tau 0} 
  - \frac{1}{2} \frac{f_{\rm R}}{2 m_B}\rho_{\rm \tau T} R_{\tau 0}
  \quad.\nonumber\\
\end{eqnarray}
\end{subequations}
$g_i$, $f_i$ are coupling constants and the indices i denote scalar (S), vector (V), tensor (T) and isovector
($\tau $). $\Phi$, $V_0$ and $R_{\tau 0}$ are the isoscalar-scalar and the zero
components of the isoscalar-vector and isovector-vector meson fields,
respectively.  The densities are defined as
corresponding local densities:
\begin{equation}
  \begin{array}{lrcl}
   \mbox{isoscalar-scalar:}
   & 
   \rho_{\rm S}(\vec{r})  
   &=&  \sum_{\alpha}\bar{\phi}_{\alpha}(\vec{r})\phi_{\alpha}(\vec{r}),
  \\[2pt]
   \mbox{isoscalar-vector:}
   & 
   \rho_{\rm 0}(\vec{r}) 
   &=& 
   \sum_{\alpha} \bar{\phi}_{\alpha}(\vec{r})\gamma_0 \phi_{\alpha}(\vec{r}),
  \\[2pt]
   \mbox{isovector-vector:}
   & 
   \rho_{\rm \tau 0}(\vec{r}) 
   &=& 
   \sum_{\alpha} 
   \bar{\phi}_{\alpha}(\vec{r})\tau_3\gamma_0\phi_{\alpha}(\vec{r}),
  \\[2pt]
   \mbox{isoscalar-tensor:}
   & 
   \rho_{\rm T}(\vec{r}) 
   &=& 
   -{\rm i} \sum_{\alpha} \vec{\nabla} \cdot (\bar{\phi}_{\alpha}(\vec{r})\vec{\alpha} \phi_{\alpha}(\vec{r})),
  \\[2pt]
   \mbox{isovector-tensor:}
   & 
   \rho_{\rm \tau T}(\vec{r})    &=& 
    -{\rm i} \sum_{\alpha}  \vec{\nabla} \cdot
   (\bar{\phi}_{\alpha}(\vec{r})\tau_3\vec{\alpha} \phi_{\alpha}(\vec{r})).
  \\[2pt]
 \end{array}
\label{eq:locdens}
\end{equation}
Now the Hamilton density will be expressed exclusively in terms of the 
densities~(\ref{eq:locdens}). To this end, 
the meson fields are eliminated by
inserting the solution of the meson field equation. This is
straightforward 
for linear coupling. We exemplify it here for the isoscalar-vector
field. The meson field equation is
$$
   \left(m_V^2-\Delta\right)V_0  =   g_V\rho_0
   -
   \frac{f_V}{2 m_B}\rho_T
   \quad.
$$
This is solved for $V_0$ by expansion in orders of $\Delta^n$ going up
to first order:
$$
   g_V V_0   =    \frac{1}{1-\Delta/m_V^2}\Big[
   \underbrace{\frac{g_V^2}{m_V^2}}_{\alpha_V} \rho_0
   -\underbrace{\frac{g_V f_V}{2 m_B m_V^2}}_{\frac{\theta_T}{2}} \rho_T\Big]
   \approx \alpha_V\rho_0 -\frac{\theta_T}{2} \rho_T+
   \underbrace{\frac{g_V^2}{m_V^4}}_{\delta_V} \Delta\rho_0 
   \quad.
$$
Reinserting that into the Hamiltonian densities ~(\ref{eq:HinitV})
yields
\begin{subequations}
\begin{equation}
  {\mathcal H}_V 
  = 
  \frac{1}{2}\alpha_V\rho_0^2 +
  \frac{1}{2}\delta_V\rho_0^2\Delta\rho_0 
  -\frac{\theta_T}{2} \rho_T  \rho_0
   +\frac{1}{2} \kappa_T\rho_T^2
\end{equation}
with recoupled strengths
\begin{eqnarray}
  \alpha_V
  &=&
  \frac{g_V^2}{m_V^2}
\\
  \delta_V
  &=&
  \frac{g_V^2}{m_V^4}
\\
  \theta_T
  &=&
  \frac{g_V f_V}{m_B m_V^2}
\\
  \kappa_T
  &=&
  \frac{f_V^2}{4 m_B^2 m_V^2}
\end{eqnarray}
\end{subequations}
The form of the recoupled strengths 
is similar for the (linear) isovector-vector term. 

The expansion is more complicated for the nonlinear isoscalar-scalar
term. It involves a combination of expansion and iteration: first, the zero-range expansion
as above, and second, an iteration of the nonlinearity. We start
from the equation determining the scalar field
\begin{eqnarray}
  g_S\Phi  =  \frac{1}{1-\Delta/m_S^2}
  \left(\alpha_S\rho_S+\tilde{b}_2\Phi^2+\tilde{b}_3\Phi^3\right),
\end{eqnarray}
where $\alpha_S$ is equal to $-g_S^2/m_S^2$ and $\tilde{b}_k$ is equal to $-b_k g_S/m_S^2$.
The meson
propagator is expanded to $\approx 1+\Delta/m_S^2$ as in the vector
case. The nonlinearity is resolved by an iteration process. We can
also obtain the $\Phi$ by using another way, for example see
Appendix \ref{app:meson}. 
We obtain the structure
\begin{subequations}
\begin{equation}
  {\mathcal H}_S= 
    \frac{1}{2}\alpha_S\rho_S^2 +
  \frac{1}{2}\delta_S\rho_S^2\Delta\rho_S
  + \frac{1}{3}\beta_S\rho_S^3
  + \frac{1}{4}\gamma_S\rho_S^4
  + \zeta_S \rho_S\Delta\rho_S + \zeta'_S \rho_S^2\Delta\rho_S + ....
\label{eq:shamilton}
\end{equation}
with coefficients 
\begin{eqnarray}
  \beta_S 
  &=& 
  -b_2 \frac{g_S^3}{m_S^6}
  \quad,
\\
  \gamma_S 
  &=& 
  (b_3 \frac{g_S^4}{m_S^8}-2b_2^2\frac{g_S^4}{m_S^{10}})
  \quad,
\\ 
  \zeta_S 
  &=& 
  -3  b_2 \frac{g_S^3}{m_S^8}
  \quad,
\\
  \zeta'_S 
  &=& 
  (5 b_3
  \frac{g_S^4}{m_S^{10}}-\frac{16}{3}b_2^2\frac{g_S^4}{m_S^{12}})
  \quad.
\end{eqnarray}
\end{subequations}

At this point, we can discuss the formal structure of the emerging
effective Hamiltonian in comparison with the point-coupling model
~(\ref{eq:HinitPC}). We see that all the terms of RMF-PC are nicely
generated by the above expansion. These are the terms with the
coefficients $\alpha_m$, $\beta_S$, $\gamma_S$, $\delta_m$, and
$\theta_m$. The expansion generates some more terms not contained in
the RMF-PC model. There is the term in $\kappa_m$, the finite range
correction for the tensor term. It can be assumed to be small because
the tensor coupling as such is already a small correction.  And there
are the many further terms generated by the expansion of the
nonlinear $\Phi$ coupling plus finite-range corrections thereof,
i.e. the terms in $\zeta_S$, $\zeta'_S$ etc. They require a more
quantitative consideration.

The mapping of the coefficients for known parametrisations can be seen
in table \ref{tab:params}. The first three columns show the effective
point-coupling parameters from the RMF-FR forces NL1 \cite{pg}, NL3
\cite{nl3} and NL-Z2 \cite{nlz2}. 
They are compared
with the parameters of two genuine point-coupling forces PC-LA
\cite{nhm} and PC-F1 \cite{buer01}.
\begin{table}[ht]
\centering
\begin{tabular}{|c|c|c|c|c|c|}
\hline Parameter  &NL3& NL1 &NL-Z2 & PC-F1& PC-LA \\\hline
$\alpha_S$ & -15.74 & -16.52  & -16.45 & -14.94& -17.55 \\
$\delta_S$ &-2.37 & -2.65  & -2.65  & -0.63& -0.64 \\\hline
$\alpha_V$ & 10.53 & 10.85  & 10.66 & 10.10& 13.34 \\
$\delta_V$ & 0.67& 0.67  & 0.68  & -0.18& -0.17 \\\hline
$\alpha_{\tau S}$ & 0 & 0  & 0 & 0& 0.029 \\
$\delta_{\tau S}$ & 0 & 0 & 0  & 0& 0 \\\hline
$\alpha_{\tau V}$ & 1.34 & 1.65  & 1.39 & 1.35& 1.27 \\
$\delta_{\tau V}$ & 0.09 & 0.11  & 0.11  & -0.06& 0 \\\hline
$\beta_S$ & 38.11 & 52.64  & 58.74 & 22.99& 3.32 \\
$\gamma_S$ & -347 & -590.77  & -706.63 & -66.76& 131.80 \\
$\zeta_S$ & 17.23 & 25.37 & 28.21 & 0& 0 \\
$\zeta'_S$ & -196.78 & -348.92  & -408.00  & 0& 0 \\\hline
$\gamma_V$ & 0 & 0 & 0  & -8.92& -100.84 \\\hline
\end{tabular}\\
\caption {Comparison of the point-coupling 
parameters between RMF-FR and RMF-PC}\label{tab:params}
\end{table}
The ``simple'' parameters $\alpha_m$ and $\delta_m$ agree
nicely amongst the various models. Comparing the $\delta_m$ between
the two models, we notice that their absolute value is smaller in
the point-coupling approach. Furthermore, as was already pointed
out in Ref. \cite{buer01}, the $\delta_m$ values for the vector
channels (both isoscalar and isovector) have a different sign
than the ones from the RMF-FR variant. This strongly indicates
that their role goes beyond the expansion of the propagators.

The parameters associated with nonlinearity show large deviations.
Moreover, one sees that the expansion of the nonlinearity within the
RMF-FR approach is slowly converging.  The expansion was done here
around $\rho_S=0$. One may hope that other expansion points, as
e.g. bulk equilibrium density, lead to better convergence. We have
checked that and find that the situation remains as bad.  A detailed
explaination can be found in appendix \ref{app:meson}. A quick check
is to take a typical value for the scalar density in bulk,
$\rho_S\approx 0.14\,{\rm fm}^{-3}$ and to multiply each term with its
power of $\rho_S$. We have then a sequence of 0.32, 0.14, 0.22 for NL1
and similar for NL-Z2.  The terms in $\zeta_S$ are not small
either. This demonstrates that the parametrization of nonlinearity is
different in its structure in both models. On the other hand, NL-Z2
and PC-F1 produce very similar results for a broad range of
observables in existing nuclei \cite{buer01}. Actual observables
explore only a small range of densities around bulk equilibrium
density, and they do not suffice to assess the underlying differences
in nonlinearity.   

To summarize: the expansion of the meson propagator of the RMF-FR into
derivative couplings in RMF-PC works fairly well for the leading
parameters. The resulting derivative couplings from RMF-FR are
different in strength and sign from those of the RMF-PC which hints
that there are other, more genuine, sources for gradient terms (quite
similar to the density functional theory for electrons \cite{GGA}).
The worst case is the expansion of the non-linearity.  The forms are
so different in RMF-FR and RMF-PC that we could not find a simple
mapping.  In order to make these differences visible in practical
applications, one needs yet to look for observables which are
sensitive to very low densities. Halo nuclei could be a promising tool
in that respect \cite{halo}.

\section{From RMF-PC to SHF}
\label{PCtoSHF}

In this section we study the nonrelativistic reduction starting
from the RMF-PC model including tensor terms and nonlinear terms in
both isoscalar-scalar and isoscalar-vector densities.  As starting
point we use the energy density of the RMF-PC model
\begin{subequations}
\label{eq:HinitPC}
\begin{eqnarray}
   {\mathcal H} 
   &=& 
   {\mathcal H}_{\rm free}
   + {\mathcal H}_{\rm S}^{\rm(PC)}
   + {\mathcal H}_{\rm V}^{\rm(PC)} 
   +{\mathcal H}_{\rm R}^{\rm(PC)}
   \quad,
\\
  {\mathcal H}_S^{\rm(PC)} 
  &=& 
  \frac{1}{2}\alpha_S\rho_S^2 +
  \frac{1}{2}\delta_S\rho_S^2\Delta\rho_S
  + \frac{1}{3}\beta_S\rho_S^3
  + \frac{1}{4}\gamma_S\rho_S^4,
\nonumber\\
{\mathcal H}_V^{\rm(PC)}
  &=& 
  \frac{1}{2}\alpha_V\rho_0^2 +
  \frac{1}{2}\delta_V\rho_0^2\Delta\rho_0 -\frac{\theta_T}{2} \rho_T
\rho_0,\nonumber\\
{\mathcal H}_{R}^{\rm(PC)}
  &=& 
  \frac{1}{2}\alpha_{\tau V}\rho_{\tau 0}^2 +
  \frac{1}{2}\delta_{\tau V}\rho_{\tau 0}^2\Delta\rho_{\tau 0}
-\frac{\theta_{\tau T}}{2} \rho_{\tau T}.
\end{eqnarray} 
\end{subequations}
This Hamiltonian contains, besides the tensor terms (isoscalar and
isovector), the isovector-vector term which appeared to be the most
important one in former investigations \cite{niko,buer01}.
The parameters $\alpha_i, \beta_i, \gamma_i, \theta_i$ are usually
determined in a $\chi^2$ adjustment to finite nuclear observables. The
tensor terms can either be put in by hand in a Hartree theory (like in
\cite{sur}) or be thought of emerging from an approximate treatment of
derivative exchange terms \cite{anto2}. 

We now want to derive the nonrelativistic limit of that energy
density following the procedure as described in \cite{pg}.  In the
first round, we consider only isoscalar fields and drop the Coulomb
interaction to keep notations simple.  Isovector contributions will be
taken into account later in the calculation of the spin-orbit terms,
because only in this sector the effect is significant.  The isoscalar
part of this Hamiltonian density leads to the stationary
Dirac-equation
\begin{subequations}
\begin{equation}
 [-i \vec{\gamma}\cdot \vec{\nabla}+m_B+S+\gamma_0 V_0
 +i \vec{\alpha}\cdot \vec{T}]\Psi_{\alpha}
 =\epsilon_{\alpha} \gamma_0 \Psi_{\alpha},  
\label{eq:Dirac}
\end{equation} 
where 
 
\begin{eqnarray}
  S
  &=&
  \alpha_S \rho_S + \delta_S \triangle  \rho_S +\beta_S
  \rho_S^2+\gamma_S \rho_S^3
  \quad,
\\
  V_0
  &=&
  \alpha_V \rho_0 + \delta_V \triangle
  \rho_0-\frac{\theta_T}{2}\rho_T
  \quad,
\\
  \vec{T}
  &=&
  -\frac{\theta_T}{2}\vec{\nabla}\rho_0
  \quad.
\end{eqnarray}
\end{subequations}
The details of the nonrelativistic expansion are given in appendix
\ref{app:details}. The nonrelativistic expansion in orders $v/c\propto
p/m$ up to $(p/m)^2$ requires, of course, small $p/m$. One also needs
to assume $\epsilon_{\alpha}\approx m_B$ which, however, is related to
the first assumption of small momenta.  The result of the expansion is
that the normal nuclear density is associated with the zeroth
component of the vector density $\rho_0$. We use henceforth the
identification $\rho_0=\rho$.  The scalar density and the vector
density are expressed in terms of this nuclear density $\rho$ and
further densities and currents as follows:
\begin{subequations}
\begin{eqnarray}
  \rho_S 
  &=& 
  \rho
  -
  2B_0^2\left(\tau-\vec{\nabla}\!\cdot\!\vec{J}+\rho  \vec{T}^2
  -
  2\vec{T}\!\cdot\!\vec{J}+  \vec{T} \!\cdot\!\vec{\nabla} \rho \right)
  \quad,
\\
  \rho_T
  &=&
  -\vec{\nabla}\cdot (B_0\vec{\nabla}\rho)
  +2\vec{\nabla}\cdot (B_0\vec{J})
  -2\vec{\nabla}\cdot (B_0 \rho \vec{T}),
\\
 B_0
 &=&
 \left[2m_B+S-V_0\right]^{-1}
  \quad.
\end{eqnarray}
The nonrelativistic densities $\rho$, $\tau$ and
$\vec {\rm J}$ are defined as 
\begin{eqnarray}
  \rho 
  &=& 
  \sum_{\alpha} W_{\alpha} \varphi^{\rm cl \dag} \varphi^{\rm cl}
  \quad, 
\\
  \tau 
  &=& 
  \sum_{\alpha} W_{\alpha}(\vec{\nabla}\varphi^{\rm cl\dag})\cdot
  (\vec{\nabla} \varphi^{\rm cl})
  \quad,
\\
  \vec{J} 
  &=&
  -\frac{i}{2} \sum_{\alpha} W_{\alpha}\left[
   \varphi^{\rm cl \dag} 
  (\vec{\nabla}\times \vec{\sigma}\varphi^{\rm cl})
  -{(\vec{\nabla}\times\vec{\sigma}\varphi^{\rm cl })}^{\dag}
   \varphi^{\rm cl} 
  \right]
  \quad,
\end{eqnarray}
\end{subequations}
where 
  $\varphi^{\rm cl}$ is the nonrelativistic single-nucleon
wavefunction.

Finally, we insert the expanded $\rho_S$ and $\rho_T$ into the
energy density (\ref{eq:HinitPC}) keeping again terms only up to second
order. This yields the nonrelativistically mapped energy density as 
\begin{eqnarray}
  {\mathcal H}^{\rm(cl)}
  &=&
  \frac{1}{2}(\alpha_S+\alpha_V)\rho^2
  +
  \frac{1}{3}\beta_S\rho^3
  +
  \frac{1}{4}\gamma_S\rho^4
  +
  \frac{1}{2}(\delta_S+\delta_V)\rho\Delta\rho
\nonumber\\
  &&
  -
  \left(
    \alpha_S\rho+\beta_S\rho^2+\gamma_S\rho^3 
  \right){\mathcal T}
\nonumber\\
  &&
  - \frac{\theta_T}{2}\rho
  \left(
    -\vec{\nabla}\cdot (B_0\vec{\nabla}\rho)
    +2\vec{\nabla}\cdot (B_0\vec{J})+\theta_T \vec{\nabla}\cdot (B_0\rho \vec{\nabla}\rho)
  \right)
  \quad,  
  \end{eqnarray}
  where
  \begin{eqnarray}
  {\mathcal T}
  &=&
  2B_0^2\left(\tau-\vec{\nabla}\cdot \vec{J}+\rho \vec{T}^2- 2\vec{T}\cdot\vec{ J}+\vec{T}\cdot \vec{\nabla}\rho \right)
\nonumber\\
  &=&
  2B_0^2\left(\tau-\vec{\nabla}\cdot \vec{J}
              +\frac{\theta_T^2}{4}\rho(\vec{\nabla}\rho)^2
              +\theta_T \vec{\nabla}\rho\cdot\vec{J}-\frac{\theta_T}{2}(\vec{\nabla}\rho)^2
  \right)
  \quad.
\end{eqnarray}

For better comparison, 
the Hamiltonian density is cast into a general form
\begin{eqnarray}
  {\mathcal H}^{\rm(cl)}
  &=&
  \frac{{\mathcal C}_1}{2}\rho^2
  +
  \frac{{\mathcal C}_2}{2}\rho\Delta\rho
  +
  {\mathcal C}_3\rho\tau
  +
  {\mathcal C}_4\rho\nabla J
  +
  \delta{\mathcal H}
\label{eq:general}
\end{eqnarray} 
where basic structures are singled out with a separate coefficient each
and the coefficients all may depend on the density, i.e.
${\mathcal C}_i={\mathcal C}_i(\rho)$. Less simple
forms are lumped together in $\delta {\mathcal H}$. This form can be
directly compared with the standard nonrelativistic mean field model,
the Skyrme-Hartree-Fock (SHF) energy functional, for a recent review see
\cite{RMPmf}. The energy functional of SHF is given in appendix
\ref{app:SHF}. It is obvious that it has the structure of the
functional (\ref{eq:general}).
Table \ref{tab:compcoeff} compares the coefficients
for the here derived nonrelativistic limit of RMF-PC with SHF.
\begin{table}
\begin{center}
\begin{tabular}{|rcll|l|}
\hline 
      && \multicolumn{1}{c}{RMF-PC basic}
       & \multicolumn{1}{c|}{RMF-PC tensor}
       & \multicolumn{1}{c|}{SHF} \\
\hline
$
 {\mathcal C}_1$ &=& $\alpha_s+\alpha_V+\frac{2}{3}\beta_S\rho+
                \frac{1}{2}\gamma_S\rho^4$
      & &
      $b_0+\frac{b_3}{3}\rho^\alpha $\\
 ${\mathcal C}_2$ &=& $\delta_S+\delta_V$  & $+\frac{\theta_T}{2}B_0-\frac{\theta_T^2}{2}B_0\rho$
      & $b_2$ \\
 ${\mathcal C}_3$ &=& $-2B_0^2\left(\alpha_S+\beta_S\rho+\gamma_S\rho^2\right) $
      & &  $b_1$ \\
 ${\mathcal C}_4$ &=& $-{\mathcal C}_3  $
      & $- \theta_T B_0$ &$ b_4$ \\
\hline
 $\delta{\mathcal H}$ &=& \multicolumn{3}{l|}{$
     \frac{\theta_T}{2}\left\{(\vec{\nabla} B_0)\cdot (\vec{\nabla}\rho)-2(\vec{\nabla} B_0)\cdot
      \vec{J}-\theta_T (\vec{\nabla} B_0 \rho)\cdot (\vec{\nabla}\rho) \right.$}\\
  &&  \multicolumn{3}{l|}{$\left.
     -2B_0^2\vec{\nabla}\rho\cdot \left(2 \vec{J}+\frac{\theta_T}{2}\rho\vec{\nabla}\rho-\vec{\nabla}\rho \right)
       \left(\alpha_s+\beta_S\rho+
                \gamma_S\rho^4
\right)
     \right\}
     $}
      \\
\hline
\end{tabular}
\end{center}
\caption{\label{tab:compcoeff} The coefficients of the energy density
(\protect\ref{eq:general}) for the nonrelativistic limit of RMF-PC
compared with the corresponding terms of SHF. The RMF-PC is
grouped in two columns.  The first column collects the terms of the
standard model. The second column adds terms stemming from tensor
coupling. Terms of the RMF-PC which do not fit into the form
(\protect\ref{eq:general}) are collected in the last two rows.
They are all related to tensor coupling.
This table shows only isoscalar terms in either model.
}
\end{table}
The table shows the similarities and the differences between the two
models. The striking similarity consists in the fact that all ${\mathcal
C}_i$ terms appear in both models. A difference appears in the relation between
$\tau$- and $\nabla J$-term. The RMF-PC without tensor coupling
predicts ${\mathcal C}_3={\mathcal C}_4$ while SHF has two separate (and
practically different) coefficients for that. It is interesting to
note that the tensor coupling in RMF-PC also allows separate
adjustment of these two terms. 
A basic difference appears with respect to the density dependence of
the coefficients. All coefficients of RMF-PC carry a more or less
involved density dependence. Only ${\mathcal C}_1$ is density dependent
for SHF and even here the form of density dependence differs.
Both models yield a very similar description of a broad range of
nuclear observables in practice. This hints that the differences
in density dependence are all somehow compensated by chosing
appropriate effective strengths. We urgently need observables for
more extreme densities to pin down the differences of the models.
All terms in the non-simple part $\delta {\mathcal H}$ come from 
the relativistic tensor coupling. It is very hard to assess their
practical importance and relative weight. One can estimate that they
are at most of the order of the $\Delta\rho$ terms. A detailed
analysis remains a task for future research.
Last not least, it ought to be mentioned that some SHF functionals
carry a term $\propto J^2$ which is not present in the above form.
It would appear in the nonrelativistic limit of RMF-PC only
in the next higher order.

To summarize: The non-relativistic limit of RMF-PC recovers the basic
structure of terms in SHF. The latter is more general in that the
kinetic and spin-orbit terms have independent parameters while they
are more or less linked in RMF. The RMF-PC, on the other hand, adds
density dependence to each one of the terms while SHF uses it only in
the leading term.

\section{Density dependent spin-orbit terms}
\label{densdepls}

This section continues the discussions of the non-relativistic limit
with a particular emphasis on the spin-orbit potential. It has
has the general structure
\begin{equation}
\vec{W}_q= b_4 \vec{\nabla} \rho + b_4' \vec{\nabla} \rho_q + c_1  \vec{J}+ c_1'  \vec{J_q},
\end {equation}
where $q=p,n$ for proton or neutron density.  The last two terms are
not found in the non-relativistic reduction. They are discarded in the
following discussion. The parameters $b_4$ and $b'_4$ are generally
density dependent. The derivation from the RMF-PC model (with
${\mathcal{H}}_{R}$ taken into account) yields in detail
\begin{eqnarray}
 b_4 &=& - \frac{A( \rho_0)}{{(2 m_q +A'( \rho_0) \rho_0 +B
\rho_q)}^2},\nonumber\\
b_4' &=& - \frac{B}{{(2 m_q +A'( \rho_0) \rho_0
+B   \rho_q)}^2},\nonumber
\end{eqnarray}
with
\begin{eqnarray}
A( \rho_0) &=& (\alpha_{\rm S}-\alpha_{\rm V}+\alpha_{\rm \tau V})+2 \beta_{\rm S}
\rho_0 + 3 (\gamma_{\rm S} -\gamma_{\rm V}) \rho_0^2,\nonumber\\
A'( \rho_0) &=& (\alpha_{\rm S}-\alpha_{\rm V}+\alpha_{\rm \tau V})+ \beta_{\rm S}
\rho_0 +  (\gamma_{\rm S} -\gamma_{\rm V}) \rho_0^2,\nonumber\\
B&=&-\alpha_{\rm \tau V}.
\end{eqnarray}
A tensor term (as it appears for example in the model of Rusnak and
Furnstahl~\cite{sur}) will induce a further correction in $b_4$ and
$b_4'$.

The spin-orbit term which is usually employed in Skyrme energy
functionals exhibits two significant differences compared to the
nonrelativistic limit of relativistic models, namely (a) the
restricted isovector dependence and (b) the lack of nonlinear terms.
An extension of the Skyrme model with an enhancement in $\vec{W}$ by
isospin contributions has already been done (SKI3-4)~\cite{ski34}.  By
introducing different isospin contributions into the spin-orbit
potential, SKI3-4 do reproduce the isotope shift of the rms radii in
the heavy Pb isotopes, but these parameter sets still yield different
shell closures in superheavy nuclei than the RMF
models~\cite{Rutz2}. On the other hand, in relativistic point-coupling
models, the role of the density dependence has proven to be important
to reproduce acceptable single particle spectra and spin-orbit
splittings \cite{buer01}. Therefore the enhancement of SKI3-4
with nonlinear terms (density dependence) might result in an improvement of
their shell structure predictions.

\begin{figure}[htb]
\centerline{\epsfig{figure=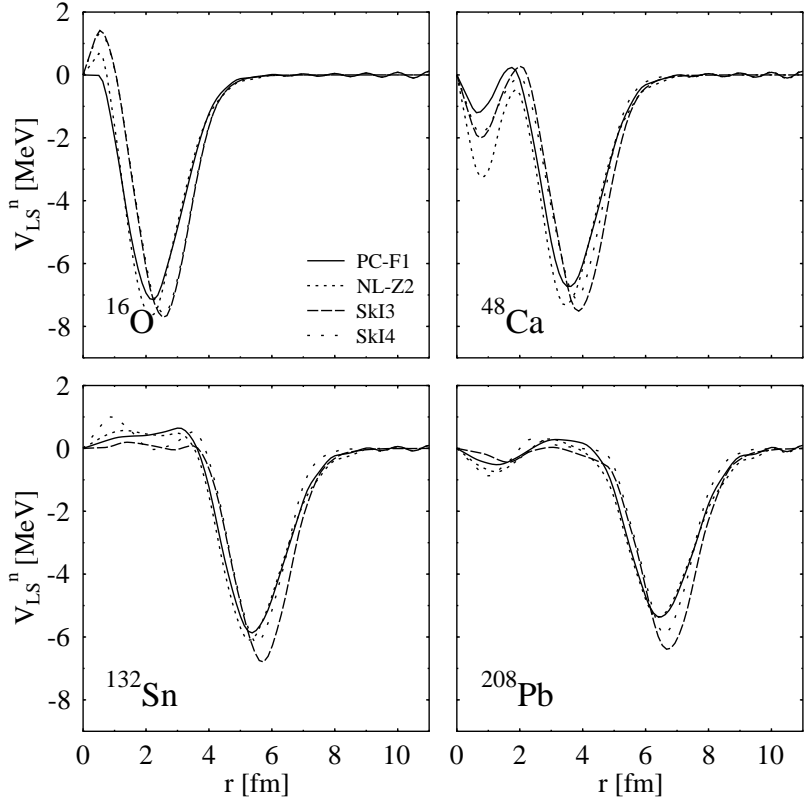,width=.8\textwidth}}
\caption{Neutron spin-orbit potential for the nuclei $^{16}$O, $^{48}$Ca, $^{132}$Sn, and $^{208}$Pb calculated with the forces as indicated.}\label{lsfig}
\end{figure}

In Fig. \ref{lsfig}, we show the neutron spin-orbit potential for the
two RMF forces PC-F1 (RMF-PC) and NL-Z2 (RMF-FR) as well as for the
Skyrme interactions SkI3 and SkI4.  (The spin-orbit potential for the
RMF was obtained from mapping the Dirac equation into an effective
Schr\"odinger equation.)  Shown are the results for the four
doubly-magic systems $^{16}$O, $^{48}$Ca, $^{132}$Sn and
$^{208}$Pb. We recognize that the potentials of the two RMF models
have a similar radial dependence, the potential of NL-Z2 being a bit
deeper in three cases. The spin-orbit potential of the Skyrme forces,
however, is both shifted to larger radii and also deeper in comparison
with the RMF results.  As could be shown in Ref. \cite{buer01}, the
RMF-PC model with PC-LA has similarly a too deep potential peaked at a
too large radius. It suffers from the same wrong trend with mass of
spin-orbit splittings as do SkI3 and SkI4 (splittings get too large
with increasing mass). The reason there are the actual values of the
nonlinear parameters, leading to a density dependence of the
mean fields that lead to this situation. This issue has been cured
with the introduction of PC-F1 which has a predictive power of
spin-orbit splittings comparable to the best RMF-FR forces. The Skyrme
force SkI3 has a spin-orbit term that mimics the isospin-dependence of
the RMF model, while SkI4 has an additional free parameter. However,
the spin-orbit potential of SkI4 lies closer to the RMF results.
A quantitative calculation using the Skyrme model not only with
isospin terms but also with density dependent terms in the spin-orbit
potential is strongly suggested. A density-dependent ansatz in the
spin-orbit potential has been introduced in Ref.~\cite{Pearson}, but
unfortunately this reference does not give a parameter set for this
type of Skyrme model. The form of the spin-orbit potential, which is
rather an {\em ad hoc} ansatz, is quite different from the one
presented here. The spin-orbit part of the energy density
which reproduces the spin-orbit potential predicted by our analysis is
given by

\begin{eqnarray}
  \varepsilon_{\rm ls}
  &\equiv &
   b_4\rho\vec{\nabla}\vec{J}
   +
   b'_4(\rho_p\vec{\nabla}\vec{J}_p+\rho_n\vec{\nabla}\vec{J}_n)
\nonumber\\
  &&
   + \frac{1}{2}(W_1 \rho + W_2
  \rho^2)(\vec{J}_n \cdot \vec{\nabla}\rho_p + \vec{J}_p \cdot
  \vec{\nabla}\rho_n + \sum_q \vec{J}_q \cdot \vec{\nabla}\rho_q).
\nonumber\\
\end{eqnarray}
The values of above parameters can be illustrated by taking the PC-F1 parameter set, using $\rho_{eq}$= $\rho_{nm}$
(PC-F1).
We obtain $b_4$= 122.93 MeV~fm$^5$,  $b'_4$=7.01 MeV~fm$^5$,
$W_1\rho_{\rm eq}/2$=36.04 MeV~fm$^5$, and $W_2(\rho_{\rm
  eq}/2)^2$=-10.26 MeV~fm$^5$.

\section{Modifications due to exchange terms}
\label{ex}

Taking into account the exchange correction in
the nonlinear terms (for more details about the calculation of these terms see Refs.~\cite{ma,burven}) , after a straight forward calculation
in the effective Hamiltonian except $C_2$, all $C_i$ 
appear to be density dependent (since the explicit form does not
provide particular insight, we omit presenting it here). If
we calculate them by using the model of Rusnak and Furnstahl~\cite{sur} with the derivative nonlinear terms
taken into account, we will
obtain $C_2$ also to be density dependent.
If we take 
into account the exchange corrections, we will have not only modifications
in $b_4$ and   $b_4'$, but  $c_1$ and   $c_1'$  also are not zero, due to the
contribution of the tensor terms which come from the Fierz
transformation~\cite{ma,burven}, these corrections are small, however.

The explicit treatment of exchange terms, due to Fierz transformations both in Iso and Dirac space, gives birth to a variety of additional isovector terms
without introducing new parameters. This is quite interesting, since
there is still a problem in the isovector channel of the RMF model: on one hand, it seems to be not flexible enough to account for isovector data, on the other hand, modern fitting strategies fail to fix additional terms corresponding to extensions in the isovector channel \cite{buer01}.
Thus, modifications governed by the treatment of exchange might be a cure.

\section{Conclusions}

We have performed a nonrelativistic reduction of the relativistic point-coupling model and compared it to both the Skyrme-Hartree-Fock model and the nonrelativistic limit of the linear RMF model with meson exchange.
The motivation was to gain more understanding about the interrelations of these different approaches, which, though looking quite different at first sight, appear to be very similar in the nonrelativistic limit.

We found that there are some significant differences in the models, namely (a) 
 there is a difference in the parametrization of the density dependence of the mean-field, (b) there is no explicit density dependence in the spin-orbit term in the SHF model.
We have written down an ansatz for a Skyrme energy functional containing these extensions which should be studied numerically in the future to see if it
can improve the spectral features of the SHF model.

A complete treatment of exchange terms in the RMF-PC model would lead to a variety of additional isoscalar and isovector terms, which, in turn, would strongly influence the nonrelativistic limit. Relativistic Hartree-Fock calculations open the possibility to a close investigation of nonrelativistic
versus relativistic kinematics and their relevance in effective models for
nuclear structure calculations.

\begin{appendix}
\section{Appendix}
\subsection{Skyrme Hartree-Fock Model (SHF)}
\label{app:SHF}

The energy density functional of standard SHF is
\begin{equation}
  \varepsilon^0 = { \varepsilon^0}_{{\rm kin}} + { \varepsilon^0}_{{\rm Sk}}(\rho, \tau, \vec{\jmath}, {\vec{J}}),  %\\
\end{equation}
the kinetic term is
\begin{equation}
{ \varepsilon^0}_{{\rm kin}} =  \, \frac{\hbar^{2}}{2m}\tau.
\end{equation}
The Skryme part reads 
\begin{eqnarray}
{ \varepsilon^0}_{{\rm SK}} & = &  \frac{b_{0}}{2} \rho^{2} - \frac{b_{0}'}{2} \sum_{q} \rho_{q}^{2}
+ \frac{b_{3}}{3} \rho^{\alpha+2} - \frac{b_{3}'}{3} \rho^{\alpha} \sum_{q} \rho_{q}^{2} \nonumber \\
& + & b_{1} (\rho\tau - \vec{\jmath}^{2}) - b_{1}'\sum_{q}(\rho_{q}\tau_{q} - \vec{\jmath}_{q}^{2})
- \frac{b_{2}}{2} \rho \Delta\rho + \frac{b_{2}'}{2} \sum_{q}\rho_{q}\Delta\rho_{q} \nonumber \\
& - & b_{4}(\rho \vec{\nabla} \cdot {\vec{J}} + \sigma \cdot (\vec{\nabla} \times \vec{\jmath})
+ \sum_{q}[\rho_{q}(\vec{\nabla} \cdot {\vec{J}}_{q}) + \sigma_{q} \cdot (\vec{\nabla} \times \vec{\jmath}_{q})]). \nonumber\\
\end{eqnarray}
Additionally, the Coulomb energy has to be added.
The densities not already defined previously are 
\begin{equation}
\vec{\jmath}_{q} = -\frac{i}{2} \sum_{i} [\psi_{i}^{\dagger}(q)\vec{\nabla}\psi_{i}(q)
- (\vec{\nabla} \psi_{i}(q))^{\dagger} \psi_{i}(q)],
\end{equation}
\begin{equation}
\vec{\sigma}_{q} = \sum_{i} \psi_{i}^{\dagger}(q) \hat{\sigma} \psi_{i}(q),
\end{equation}
but these densities are zero in spherically symmetric systems due to  time-reversal invariance.

\subsection{Walecka Model (RMF-FR) }
\label{app:RMF}
The Lagrangian density for RMF-FR is
\begin{equation}
{\mathcal{L}} = {\mathcal{L}}^{\rm free}_{\rm nucleon} + {\mathcal{L}}^{\rm free}_{\rm meson} + 
    {\mathcal{L}}^{\rm lin}_{\rm coupl} + {\mathcal{L}}^{\rm nonlin}_{\rm coupl},   
\end{equation}
where
\begin{equation}
{\mathcal{L}}^{\rm free}_{\rm nucleon} = \bar{\psi} (i\gamma_{\mu}\partial^{\mu} - m_{\rm B} ) \psi, 
\end{equation}
\begin{eqnarray}
{\mathcal{L}}^{\rm free}_{\rm meson} & = & \frac{1}{2}(\partial_{\mu}\Phi\partial^{\mu}\Phi - m_{s}^{2} \Phi^{2}) \nonumber \\
                       & - & \frac{1}{2}(\frac{1}{2} G_{\mu \nu} G^{\mu \nu} - m_{v}^{2} V_{\mu} V^{\mu}) \nonumber \\
                      &  -  & \frac{1}{2}(\frac{1}{2} \vec{B}_{\mu \nu} \cdot \vec{B}^{\mu \nu} - m_{r} 
^{2} \vec{R}_{\mu} \cdot \vec{R}^{\mu}) \nonumber \\
                     &   - & \frac{1}{4} F_{\mu \nu} F^{\mu \nu} ,\nonumber \\
\end{eqnarray}

\begin{eqnarray}
{\mathcal{L}}^{\rm lin}_{\rm coupl} &=& - g_{S} \Phi \bar{\psi} \psi - g_{V} V_{\mu} \bar{\psi} \gamma^{\mu} \psi 
                          - g_{R} \vec{R}_{\mu} \cdot \bar{\psi}
                          \vec{\tau} \gamma^{\mu} \psi \\ 
&-&\frac{i f_{V}}{2 m_B}\partial_{\nu} V_{\mu} \bar{\psi} \gamma^{\mu}  \gamma^{\nu} \psi  
-\frac{i f_{R}}{4 m_B}\partial_{\nu} \vec{R}_{\mu} \bar{\psi} \vec{\tau}
\gamma^{\mu}  \gamma^{\nu} \psi\\ &-&
                            e A_{\mu} \bar{\psi} \frac{1+\tau_{3}}{2} \gamma^{\mu} \psi,\nonumber\\
\end{eqnarray}
and
\begin{equation}
 {\mathcal{L}}^{\rm nonlin}_{\rm coupl} = - \frac{1}{3} b_{2} \Phi^{3}
                   - \frac{1}{4} b_{3} \Phi^{4}.%\nonumber\\
\end{equation}

\subsection{Nonlinear Scalar Meson Equation}
\label{app:meson}
Here, we give an alternative approach ~\cite{Bir} to arrive at
Eq.~(\ref{eq:shamilton}). In this approach it becomes transparent that the convergence of the 
nonlinear expansion of the scalar density depends solely on $g_s$ and $b_i$ and
not on the expansion value of the scalar density. The equation for the scalar meson can be written as 
\begin{eqnarray}
\label{eq:mes}
(\nabla - m_s^2)\Phi= g_s \rho_s + b_2 \Phi^2 +b_3 \Phi^3.  
\end{eqnarray}
One way to solve this equation is by using an iteration procedure as follows:
we start by choosing an initial $\Phi$:
\begin{eqnarray} 
\Phi_0(r_0)=-gs \int D(|\vec{r_0}-\vec{r_1}|)\rho_s(\vec{r_1}) d^3r_1,
\end{eqnarray}
where $ D(|\vec{r_0}-\vec{r_1}|)$ is the Greens function satisfying $(\nabla -
m_s^2)\Phi_0 = g_s \rho_s$, and substitute this $\Phi$ into
\begin{eqnarray}
\label{eq:mes1}
(\nabla - m_s^2)\Phi_n= g_s \rho_s + b_2 \Phi_{n-1}^2 +b_3 \Phi_{n-1}^3.
\end{eqnarray}
We can say that the iteration procedure terminates if $\Phi_k$=$\Phi_{k+1}$.
The result is
\begin{eqnarray}
\label{eq:mes2}
\Phi (r )&=&-gs \int D(|\vec{r}-\vec{r_1}|)\rho_s(\vec{r_1})
d^3r_1\nonumber\\&+& \int \int
f_2(\vec{r},\vec{r}_1,\vec{r}_2)\rho_s(\vec{r_1})\rho_s(\vec{r_2})d^3r_1
d^3r_2\nonumber\\&+& \int
\int\int
f_3(\vec{r},\vec{r}_1,\vec{r}_2,\vec{r}_3)\rho_s(\vec{r_1})\rho_s(\vec{r_2})
\rho_s(\vec{r_3}) d^3r_1 d^3r_2
d^3r_3\nonumber\\&+& \cdot \cdot \cdot
\end{eqnarray}
where 
\begin{eqnarray}
f_2(\vec{r},\vec{r}_1,\vec{r}_2)= -g_s^2 b_2  \int D(|\vec{r}-\vec{r}'|)D(|\vec{r}'-\vec{r_1}|)D(|\vec{r}'-\vec{r_2}|)d^3r'
\end{eqnarray}
\begin{eqnarray}
f_3(\vec{r},\vec{r}_1,\vec{r}_2,\vec{r}_3)&=& g_s^3 b_3  \int
D(|\vec{r}-\vec{r}'|)D(|\vec{r}'-\vec{r_1}|)D(|\vec{r}'-\vec{r_2}|)D(|\vec{r}'-\vec{r_3}|)d^3r'\nonumber\\&-&g_s^3
b_2^2 \int \int D(|\vec{r}-\vec{r}''|)D(|\vec{r}''-\vec{r}'|)D(|\vec{r}''-\vec{r_1}|)D(|\vec{r}'-\vec{r_2}|)\nonumber\\
&\cdot & D(|\vec{r}'-\vec{r_3}|)d^3r'd^3r'',
\end{eqnarray}
... is the contribution from terms with $\rho_s$ more than three.
It is clear  from Eq. (\ref{eq:mes2}) that the fast convergence in the
nonlinearity ( in the power of $\rho_s$) can only be obtained if the coupling
constants $g_s$ and $b_i$ are small, basically independent of the point of expansion.

Because $D(|\vec{r}''-\vec{r}'|)$ is a distribution function, we can
expand it into a delta function and its derivative:
\begin{eqnarray}
D(\vec{r},\vec{r}')= \frac{1}{m_s^2} \delta^3(\vec{r}-\vec{r}')+\frac{\nabla}{m_s^4} \delta^3(\vec{r}-\vec{r}')+...  
\end{eqnarray}
If we choose to expand only up to the second term, insert it into
$\Phi$ and insert $\Phi$ into the scalar Hamiltonian, we 
obtain the same result as Eq.~(\ref{eq:shamilton}). This
delta expansion, if integrated with density, has a connection with the Taylor
expansion of the density. It can be easily understood from an artificial
1 dimensional (1D) illustration as follows:
in 1D, the propagator can be written as

\begin{eqnarray}
D(x)&=& - \frac{g_s}{m_s^2} \delta(x) +\frac{g_s}{m_s^4} \delta'(x)+..+
g_s \frac{(-1)^{n+1}}{{m_s}^{2(n+1)}} \delta^{(n)}(x),\nonumber\\
&=& f_0 \delta(x) + f_1 \delta'(x)+..+
f_n \delta^{(n)}(x)  
\end{eqnarray}
where $\delta^{(n)}(x)$= $d^n/dx^n \delta(x)$.
Then $\Phi_0 (0)$ is
\begin{eqnarray}
\Phi_0 (0)= \int_{- \infty}^{\infty} \rho(x)D(x)dx=f_0
\rho(0)-f_1\rho'(0)+...+(-1)^n \rho^{(n)}(0).\nonumber\\
\end{eqnarray}
Now if we expand $\rho(x)$ around x=0 in a Taylor series as
\begin{eqnarray}
\rho(x)= \rho(0)+\rho'(0)x +...+ \frac{1}{n !}  \rho^{(n)}(0),
\end{eqnarray}
we obtain
\begin{eqnarray}
\Phi_0 (0)= \rho(0) \int_{- \infty}^{ \infty} D (x)dx +\rho'(0) \int_{-
  \infty}^{ \infty} x D(x)dx +...+ \rho^{(n)}(0) \int_{- \infty}^{ \infty}
\frac{1}{n !} x^n D(x)dx.\nonumber\\
\end{eqnarray}
If we compare both $\Phi_0$, we have an alternate representation of D(x):
\begin{eqnarray}
D(x)&=&\sum_0^{\infty}\frac{(-1)^n}{n !}\delta^{(n)}(x)  \int_{- \infty}^{ \infty}
\frac{1}{n !} {x'}^n D(x')dx' 
\end{eqnarray}
It is clear that the choice of proper origin for the expansion in $x$ has an effect on how far we need to expand
$\rho (x)$ to obtain a good approximation.
If we for example find  x=0 as a good position, so that
$\rho(x)\approx \rho(0)+\rho'(0)x$, it means that for every $f_n$, n$>$1
gives contribution zero. The last equation tells us that it is nothing else than
$D(x)$ $\approx$ - $\frac{g_s}{m_s^2} \delta(x)$ +$\frac{g_s}{m_s^4}
\delta'(x)$. We suspect a similar behavior to happen in the real word (3D).
\subsection{Details of the nonrelativistic reduction}
\label{app:details}

This appendix provides a brief outline of 
the nonrelativistic expansion of the scalar and tensor densities.
For simplicity, the wavefunctions are used without the index 
for the state $\alpha$.

The Dirac equation (\ref{eq:Dirac}) decomposes into
\begin{subequations}
\begin{eqnarray}
  (m-\epsilon+S+V_0)\varphi^{(up)}+\sigma\!\cdot\!(p+iT)\varphi^{(dw)} 
  &=&
  0
  \quad,\\
  (m+\epsilon+S-V_0)\varphi^{(dw)}-\sigma\!\cdot\!(p-iT)\varphi^{(up)} 
  &=&
  0
  \quad.
\end{eqnarray}
\end{subequations}
The scalar and vector densities are
\begin{subequations}
\begin{eqnarray}
  \rho_s
  &=&
  \left|\varphi^{(up)}\right|^2
  -
  \left|\varphi^{(dw)}\right|^2
  \quad,\\
  \rho_0
  &=&
  \left|\varphi^{(up)}\right|^2
  +
  \left|\varphi^{(dw)}\right|^2
  \quad.
\end{eqnarray}
\end{subequations}
The ``normal'' (baryon) density is the vector density $\rho_0$. 
In the following, we eliminate the lower component thereby carrying
forth only terms up to second order in $p/m$. 

The lower component can be expressed through the upper component
\begin{subequations}
\begin{eqnarray}
  \varphi^{(dw)}
  &=&
  B_0\sigma\!\cdot\!(p-iT)\varphi^{(up)}
  \quad,\\
  B_0
  &=&
  \frac{1}{m+\epsilon+S-V_0}
  \approx
  \frac{1}{2m+S-V_0}
  \quad.
\end{eqnarray}
The upper component is not yet the nonrelativistic wavefunction
because as such it is not normalized. The classical wavefunction
is introduced through
\begin{eqnarray}
  \varphi^{(up)}
  &=&
  \hat{I}^{-1/2}  \varphi^{(cl)}
  \quad,\\
  \hat{I}
  &=&
  1+\sigma\!\cdot\!(p+iT)B_0^2\sigma\!\cdot\!(p-iT)
  \quad.
\end{eqnarray}
\end{subequations}
This yields the desired result for the vector density
\begin{equation}
  \rho_0  =   \left|\varphi^{(cl)}\right|^2
  \quad.
\end{equation}
More involved terms appear for the scalar density: 
\begin{eqnarray}
  \rho_s
  &=&
  \left|\varphi^{(up)}\right|^2
  -
  \left|\varphi^{(dw)}\right|^2
  \quad,
\nonumber\\
  &=&
  {\varphi^{(cl)}}^+\hat{I}^{-1}\varphi^{(cl)}
  -
  {\varphi^{(cl)}}^+\hat{I}^{-1/2}
  \sigma\!\cdot\!(p+iT)B_0^2\sigma\!\cdot\!(p-iT)
  \hat{I}^{-1/2}\varphi^{(cl)}
\nonumber\\
  &=&
  \rho_0
  -
  2
  {\varphi^{(cl)}}^+
  \sigma\!\cdot\!(p+iT)B_0^2\sigma\!\cdot\!(p-iT)
  \varphi^{(cl)}
\nonumber\\
  &=&
  \rho_0
  -
  2B_0^2\left(\tau-\nabla\!\cdot\!J+\rho_0T^2-2T\!\cdot\!J+T \cdot\ \nabla\rho \right)
  \quad.
\end{eqnarray}
The tensor density is expanded as
\begin{eqnarray}
  \rho_T
  &=&
  \nabla\varphi^+
  \left(\begin{array}{cc}
    0 & -i\sigma \\ i\sigma & 0
  \end{array}\right)
  \varphi
\nonumber\\
  &=&
  -\nabla{\varphi^{(up)}}^+\imath\sigma B_0\sigma\!\cdot\!(p-iT)\varphi^{(up)}
  +\nabla{\varphi^{(up)}}^+\!\cdot\!\sigma(p+iT)B_0\imath\sigma\varphi^{(up)}
\nonumber\\
  &\approx&
  -\nabla(B_0\nabla\rho_0)
  +2\nabla(B_0J)
  - \nabla(B_0 T\rho_0)
  \quad.
\end{eqnarray}
\end{appendix}

\section*{Acknowledgements}

The authors would like to thank M.~Bender for stimulating
discussions. A.S. gratefully acknowledges financial support from the DAAD.
This work was supported in part by  Bun\-des\-ministerium f\"ur 
Bildung und Forschung (BMBF), Project No.\ 06 ER 808.

\end{document}